\documentclass[10pt,journal,compsoc]{IEEEtran}
\usepackage{amsmath,amssymb,amsfonts}
\usepackage{algorithmic}
\usepackage{graphicx}
\usepackage{textcomp}
\usepackage{wrapfig}
\usepackage{subfigure}
\usepackage{makecell}
\usepackage{colortbl}
\usepackage{url}
\usepackage{bm}

\ifCLASSOPTIONcompsoc
  \usepackage[nocompress]{cite}
\else
  \usepackage{cite}
\fi

\ifCLASSINFOpdf
\else
\fi

\begin{document}

\title{Characteristics of Interference in Licensed and Unlicensed Bands for Intelligent Spectrum Management}
 \author{ Zhuoran Su,~\IEEEmembership{Graduate Student Member,~IEEE} and
            Kaveh Pahlavan ,~\IEEEmembership{Life Fellow, IEEE}
\IEEEcompsocitemizethanks{\IEEEcompsocthanksitem The  work  presented  in  this  paper is submitted as ECE538 Final Project.\protect\\
}}

\markboth{Journal of \LaTeX\ Class Files,~Vol.~14, No.~8, August~2015}%
{Shell \MakeLowercase{\textit{et al.}}: Bare Demo of IEEEtran.cls for Computer Society Journals}

\IEEEtitleabstractindextext{%
\begin{abstract}
The exponential growth of IoT devices and the demand of smart devices for higher data rates has heightened the need for sharing and managing spectrum resources in cellular 5G/6G operating in licensed bands and Wi-Fi technologies operating in unlicensed bands. Intelligent spectrum management has emerged as a key concept in dynamic spectrum allocation. To understand the interference existing in the spectrum, researchers usually monitor the interference in a fixed location and either focus on the cellular band or Wi-Fi band. In this study, we conduct experiments for collecting real-time spectrum data in indoor and outdoor environments with a mobile receiver, the spectrum analyzer. For outdoor, we mount the spectrum analyzer on a car seat and drive on the selected route in an urban area. We put the analyzer on a cart and moved it around in the laboratory for indoor. The frequency of interest in this study is 1.9 – 2.5 GHz, including both licensed and unlicensed bands. Temporal and frequency domain behavior is compared between licensed and unlicensed bands. We first normalize and binarize the data with a threshold. Then we calculate the spectrum occupancy by counting how many consecutive ones. Based on our observation, the spectrum occupancy of the outdoor environment is more remarkable than the indoor environment. The interference in the licensed band shows more variations in the frequency domain than that in the unlicensed band. This study provides a better understanding of the interference behavior for different environments and frequency bands.
\end{abstract}

\begin{IEEEkeywords}
Characteristics of Interference modeling, licensed band, unlicensed band, spectrum management
\end{IEEEkeywords}}

\maketitle

\IEEEdisplaynontitleabstractindextext

\IEEEpeerreviewmaketitle

\IEEEraisesectionheading{\section{Introduction}\label{sec:introduction}}

\IEEEPARstart{R}ecently, the intelligent spectrum management has received more attention than past few years since it shows the ability to efficiently allocate the scarce frequency resource. It is one of the key techniques for next-generation communication systems to liberate the spectrum\cite{pahlavan2021understanding}. However, the wireless service provider and the airlines initiated a discussion about whether the 5G can affect the safety of aircrafts\cite{news2022}. The cornerstone of intelligent spectrum management, interference monitoring and modeling, can be a solution to forecast the interference and dynamically allocate the spectrum without affecting the safety of other systems. 
The CRN (Cognitive radio network) has been well researched since it plays an essential role in spectrum sharing\cite{okegbile2021stochastic}. It requires accurate interference modeling to protect the user’s communication. Related work in\cite{yun2021intelligent} proposed an intelligent dynamic spectrum resource management mechanism that can coexist with other CRNs. The stochastic geometry-based models get rid of extensive Monte Carlo simulations and provide methods for random spatial patterns. This kind of model treats the locations of the base stations as points distributed\cite{wang2016stochastic}, and they are widely used in analyzing the performance of CRN. Using stochastic geometry tools, authors in\cite{lu2021stochastic} presented a comprehensive spatial-temporal analysis of large-scale communications systems.

\begin{figure*}[t!]
\centering
\hfill
\subfigure[]{
  \begin{minipage}[b]{0.48\textwidth}
    \centering
    \includegraphics[width=0.75\textwidth,scale=0.3]{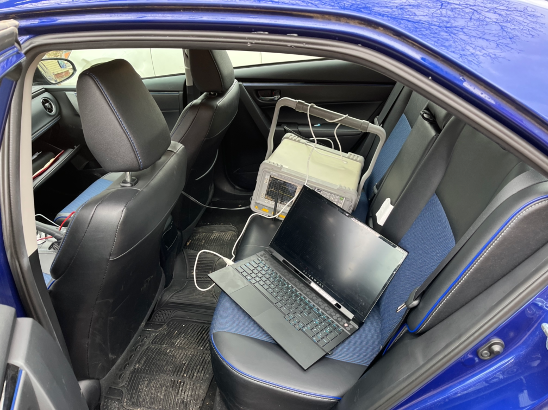}
    \label{fig:1_1}
  \end{minipage}
}
\hfill
\subfigure[]{
  \begin{minipage}[b]{0.48\textwidth}
    \centering
    \includegraphics[width=0.75\textwidth,scale=0.3]{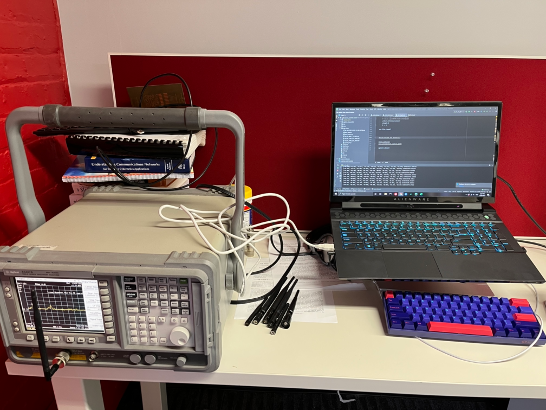}
    \label{fig:1_2}
  \end{minipage}
}

\caption{Measurement scenarios in our dataset, (a) Outdoor (b) Indoor}
\end{figure*}

\begin{figure*}[t!]
\begin{minipage}[b]{\textwidth}
    \centering
    \includegraphics[width=\textwidth]{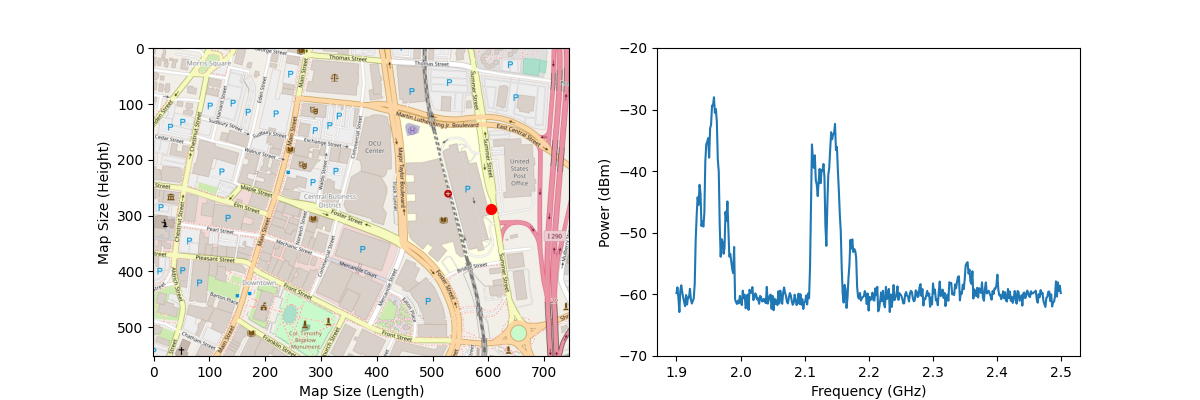}
  \end{minipage}
  \caption{Trajectory and Measurement \label{fig:2}}
\end{figure*}

Related works successfully establish models for Wi-Fi bands to determine whether the channel is busy. \cite{hou2021modeling} measured the 2.4 GHz and 5GHz Wi-Fi bands and proves the current channel allocation is not efficient enough. 7 different distributions are evaluated using Kolmogorov-Smirnov (KS) distance, Kullback-Leibler (KL) divergence, and Bhattacharyya distance to compare the predictability. Authors in \cite{hou2021modeling} monitored the 2.4 GHz and 5 GHz bands in a railway station with fixed location receivers. Spectrogram plays a vital role in the analysis of temporal features. To exploit the spectrogram using Deep Learning algorithms, \cite{bhatti2021shared} proposes the Q-spectrogram and shows it is better for CNN than the traditional spectrogram using experimental data. The Deep Learning model trained on Q-spectrogram offers 99\% accuracy in estimating the Wi-Fi traffic load (five density levels). Instead of using the locations of base stations, \cite{al2018free} monitors 915-928 MHz ISM-band in Melbourne, Australia. The authors stated that the normalized histogram follows a log-normal distribution. The duty cycles are calculated across all the frequency bins as the key parameters for evaluating occupancy. The definition of whether a band is busy varies in the literature, But for Wi-Fi bands, the binarization of data can simplify the modeling. \cite{al2018free} and \cite{hou2021modeling} converted the data into binary patterns and counted the length of the continuous busy and idle durations with certain resolutions. The threshold for binarization also varies in the literature since it depends on the design purpose of the application. In this work, we follow a similar approach to analyze the interference behavior in 1.9-2.5 GHz. Most of the related works focus only on the unlicensed bands without considering the licensed band. The problem is if we can use the same method to model the interference in licensed bands and if we can monitor and predict the occupancy in all types of frequency bands.  
To explore the possibility of allocating both licensed and unlicensed bands, we collected interference measurements in the spectrum in an urban area in Worcester, MA, and in a laboratory building of WPI. We proposed an interference model based on both temporal and spatial information. During the data collection, we added the GPS tag to each measurement to prove interference in the licensed band is highly related to location information. The rest of this paper is organized as follows; We first introduce the importance of spectrum monitoring and interference modeling. In part \ref{sec:2}, we present the scenario and equipment in the data collection phase and show the result of spectrum monitoring. In part \ref{sec:3}, we construct the interference model and analyze the interference behavior based on the real data. In part 4, we state the importance of interference modeling in intelligent spectrum management and present the Contribution of this study.

\section{The Measurement Scenarios and Data Analysis}
\label{sec:2}
In this section, we will introduce the measurement scenarios and dataset size and structure. The equipment used in the study is Agilent E4407B ESA-E Spectrum Analyzer, which can measure 9kHz to 26.5 GHz with 0.4 dB overall amplitude accuracy. The frequency of interest is 1.9 GHz to 2.5 GHz, including the LTE band and 2.4 GHz ISM band. Therefore, we are able to compare the interference for different types of bands. Fig.\ref{fig:1_1} shows the outdoor measurement scenario. We put the spectrum analyzer on the back seat of the car and connected it with a laptop using a GPIB cable. Benefit from PyVISA, we can easily control the equipment and retrieve data from it. The GPS receiver is also connected to the laptop to label the data with GPS coordinates. For the indoor environment shown in Fig.\ref{fig:1_2}, we put the spectrum analyzer on a cart and move it around the third floor of Atwater Kent laboratory of WPI. We do not use GPS in this scenario since the GPS receiver performs poorly in the indoor environment, which can have an error up to tens of meters.

Fig.\ref{fig:2} shows a test drive on the selected route. We select an area near Worcester Common, Worcester, MA, an urban region with a large population. There are at least 5 cellular towers around. Fig. 2 left is the map of the selected area, and the trajectory is marked as a solid red dot. The relative location is calculated by GPS coordinates. Fig. 2 right is the spectrum measurement corresponding to this location. The frequency range is 1.9 GHz to 2.5 GHz, and the amplitude is from -20 dBm to -70 dBm. Each test drive is about 10 - 15 minutes, depending on the traffic situation. We can collect about 700 – 1000 measurements during the drive. Each measurement consists of 401 amplitude readings representing the 600 MHz frequency span and the 2.2 GHz center frequency. The omnidirectional antenna is used in this study. Since the noise level changes over time for different measurements, we normalized the receiver power to make fair comparisons.

\begin{figure*}[t!]
\centering
\hfill
\subfigure[]{
  \begin{minipage}[b]{0.48\textwidth}
    \centering
    \includegraphics[width=\textwidth]{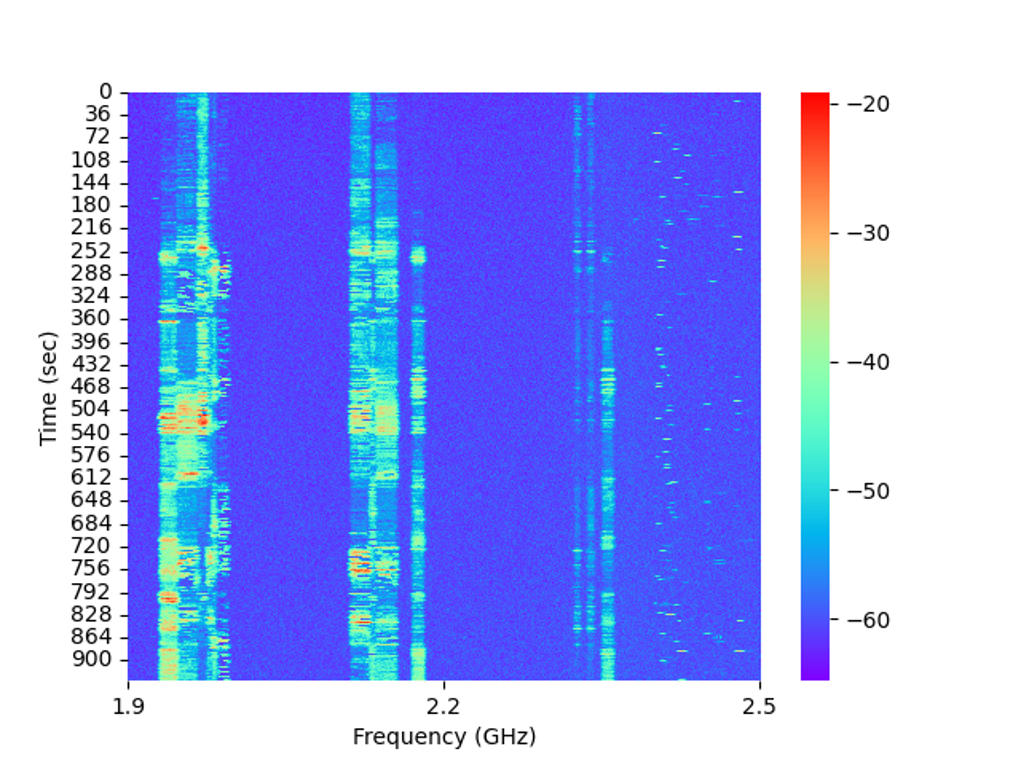}
    \label{fig:3_1}
  \end{minipage}
}
\hfill
\subfigure[]{
  \begin{minipage}[b]{0.48\textwidth}
    \centering
    \includegraphics[width=\textwidth]{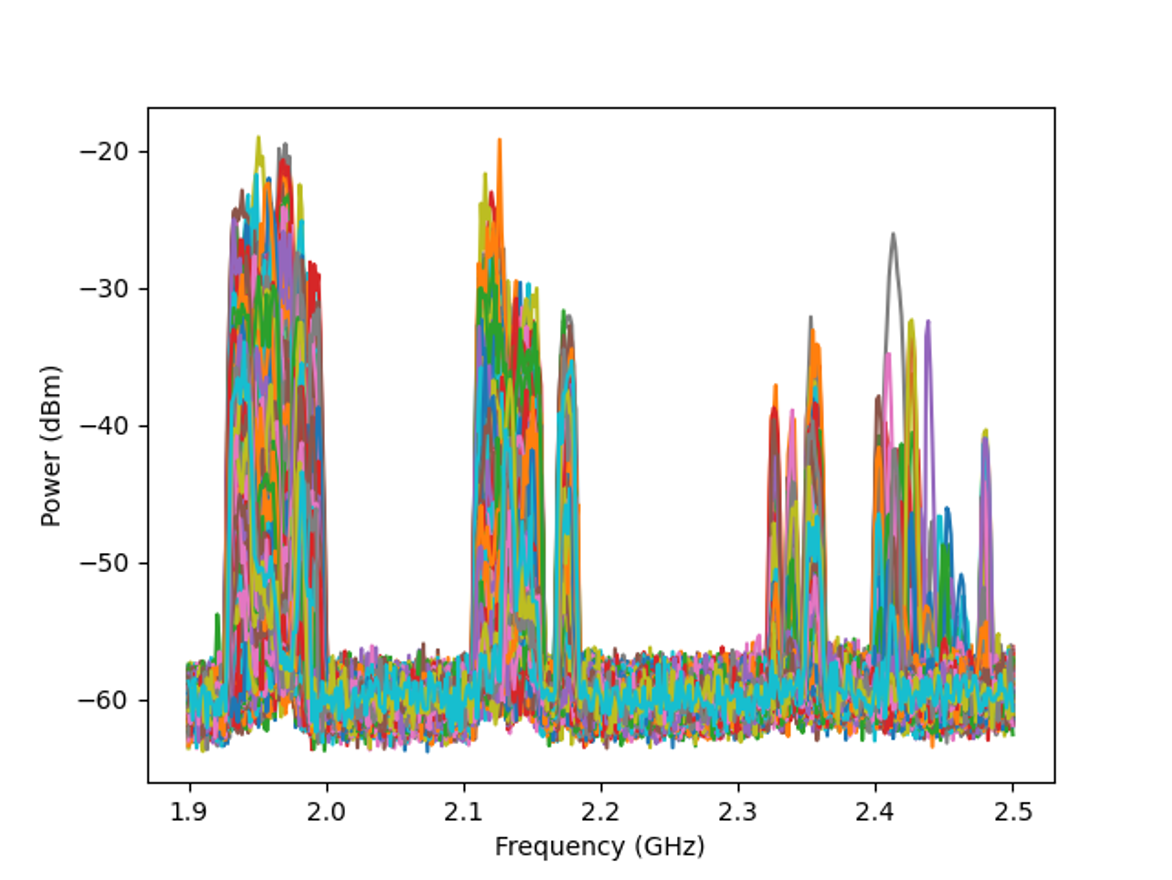}
    \label{fig:3_2}
  \end{minipage}
}

\caption{(a) Spectrogram of One Test Drive (b) Congested Plot}
\end{figure*}

Without processing the collected data, we look into the raw data for an overview of temporal and spatial behavior of the interference. To analyze the difference between licensed and unlicensed bands, we plot the spectrogram of the data. Fig 3 is one set of data, including a complete test drive. Fig.\ref{fig:3_1} is the spectrogram of the measurement starting at 0 sec and ending at 950 sec, and the power is between -20 dBm and -70 dBm. In this figure, different times also represent other locations. For licensed bands, we observe higher interference at locations closer to the cell tower (around 504 sec) than on the open road (0 – 216 sec). 1.9 GHz – 2.0 GHz has a similar spatial pattern as 2.1 – 2.2 GHz since the variations in power levels are almost the same. For the 2.4 GHz unlicensed band, the interference is discontinuous, and the power level is significantly lower than the licensed bands. Fig.\ref{fig:3_2} is the congested plot of the data showing the interference between 1.9 GHz and 2.5GHz. 2.0 GHz – 2.1 GHz and 2.2 -2.3 GHz seems to be inactive bands during the experiment, which shows the potential of spectrum sharing applications. 

\begin{figure*}[t!]
\centering
\hfill
\subfigure[]{
  \begin{minipage}[b]{0.3\textwidth}
    \centering
    \includegraphics[width=\textwidth]{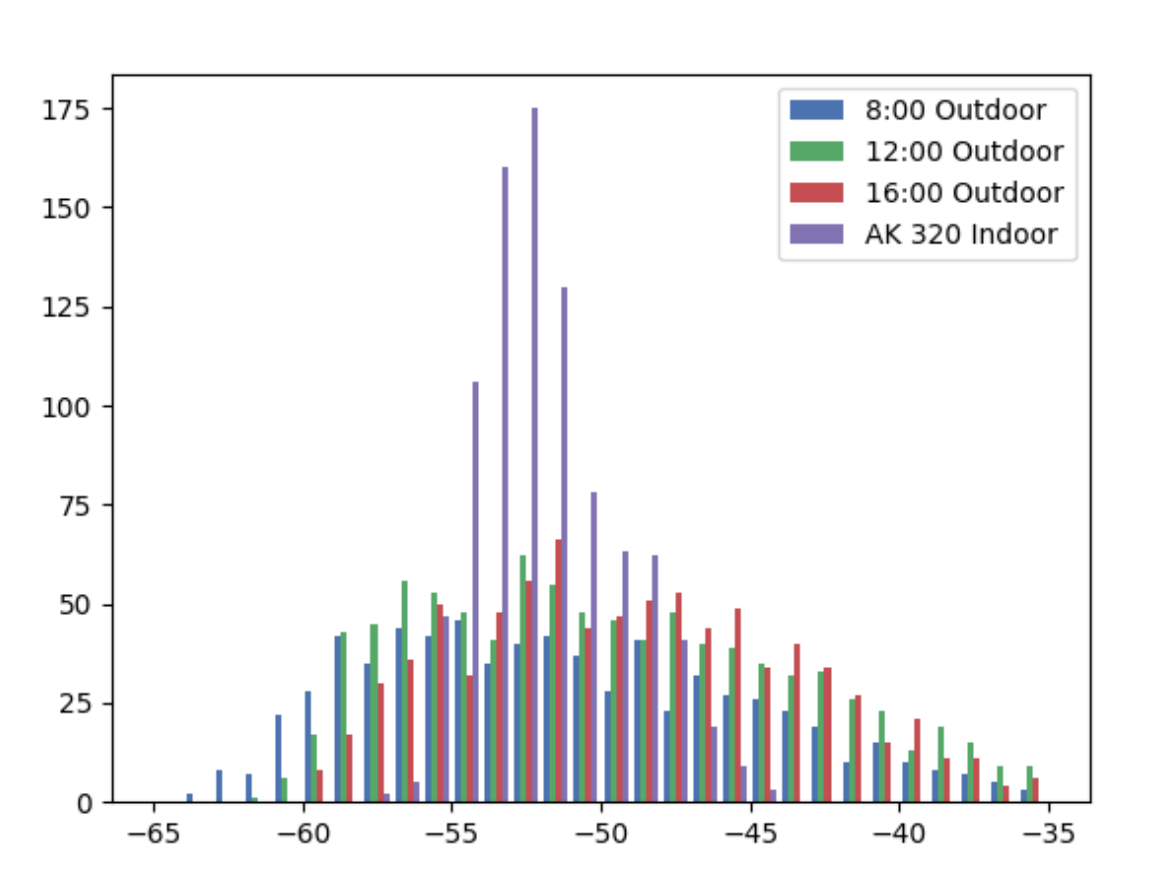}
    \label{fig:new_1}
  \end{minipage}
}
\hfill
\subfigure[]{
  \begin{minipage}[b]{0.3\textwidth}
    \centering
    \includegraphics[width=\textwidth]{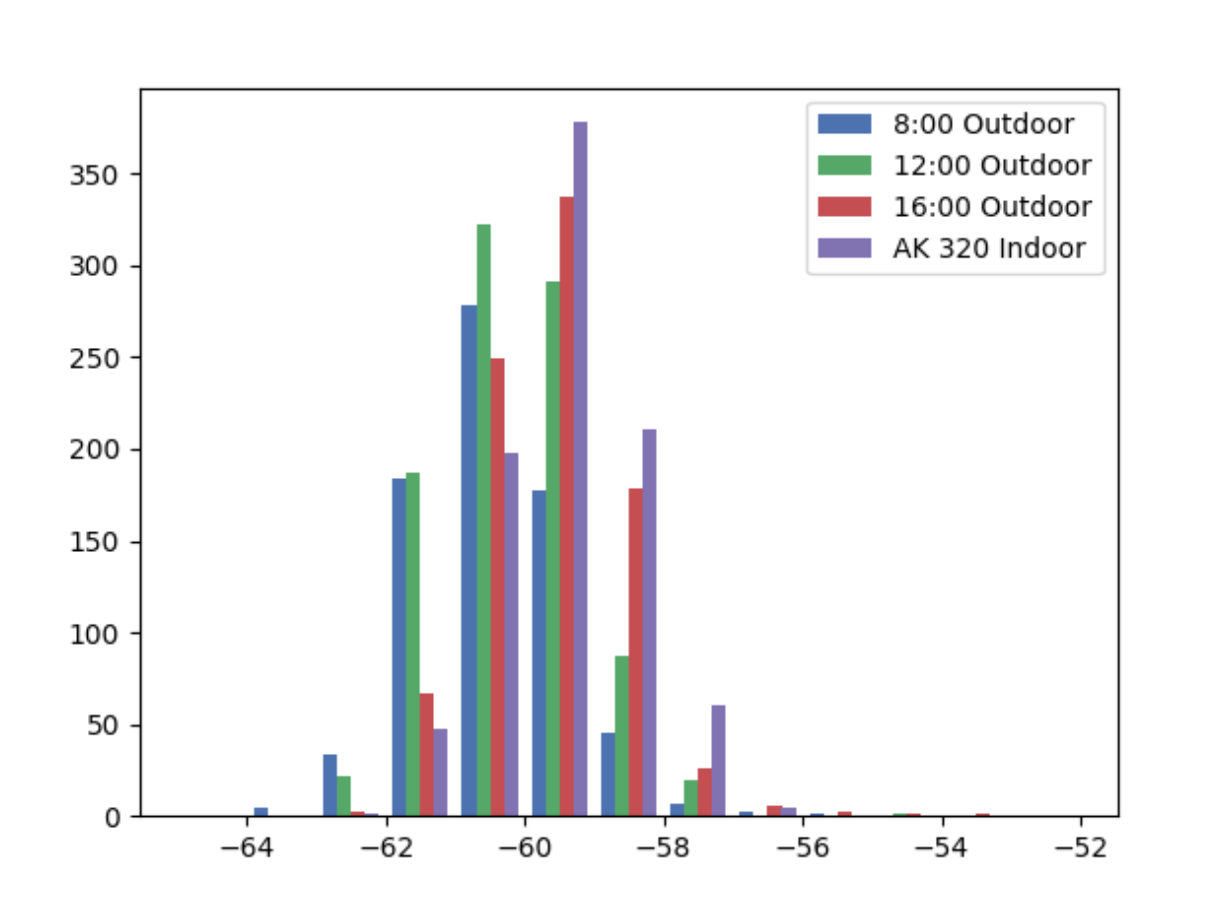}
    \label{fig:new_2}
  \end{minipage}
}
\hfill
\subfigure[]{
  \begin{minipage}[b]{0.3\textwidth}
    \centering
    \includegraphics[width=\textwidth]{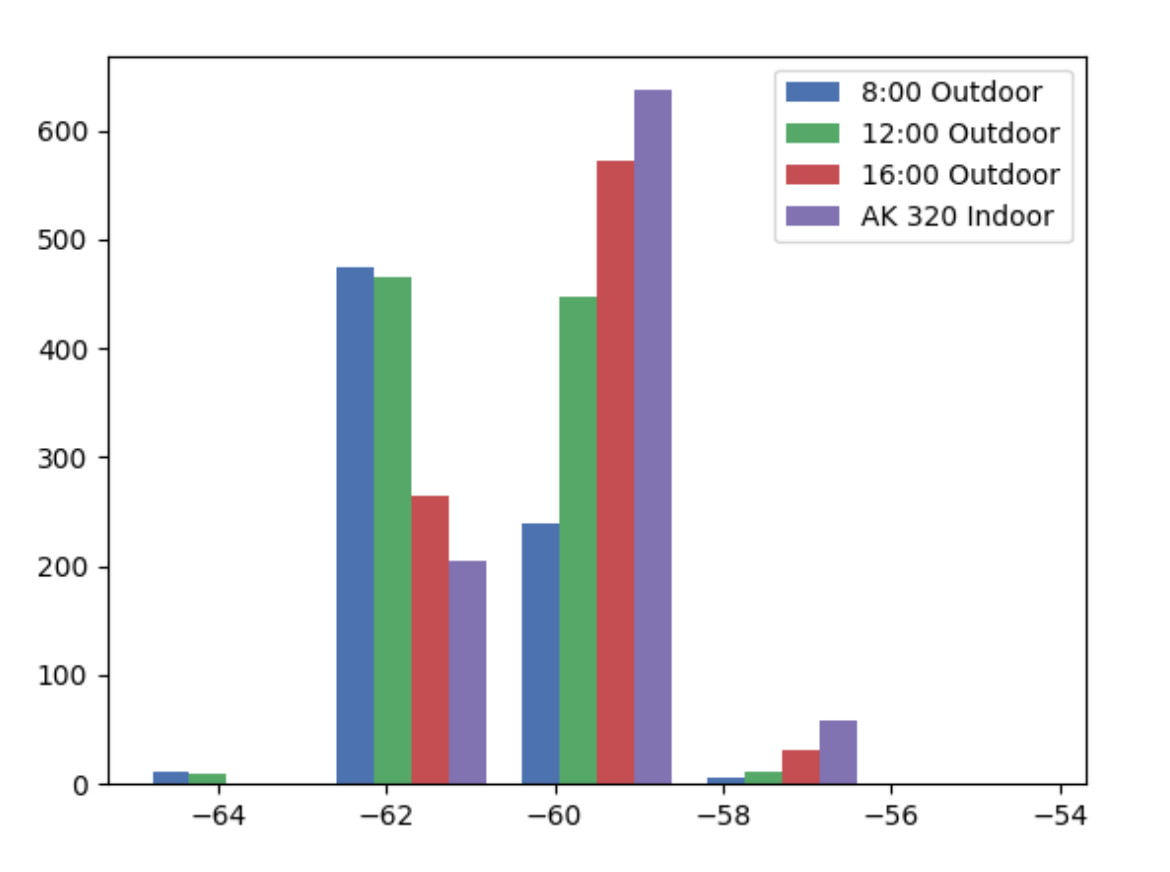}
    \label{fig:new_3}
  \end{minipage}
}

\caption{Histograms of 4 Datasets in Different Bands: (a) Licensed (b) Unlicensed (c) Inactive}
\end{figure*}

\begin{figure}[t!]
\centering
\hfill
  \begin{minipage}[b]{0.48\textwidth}
    \centering
    \includegraphics[width=0.75\textwidth,scale=0.3]{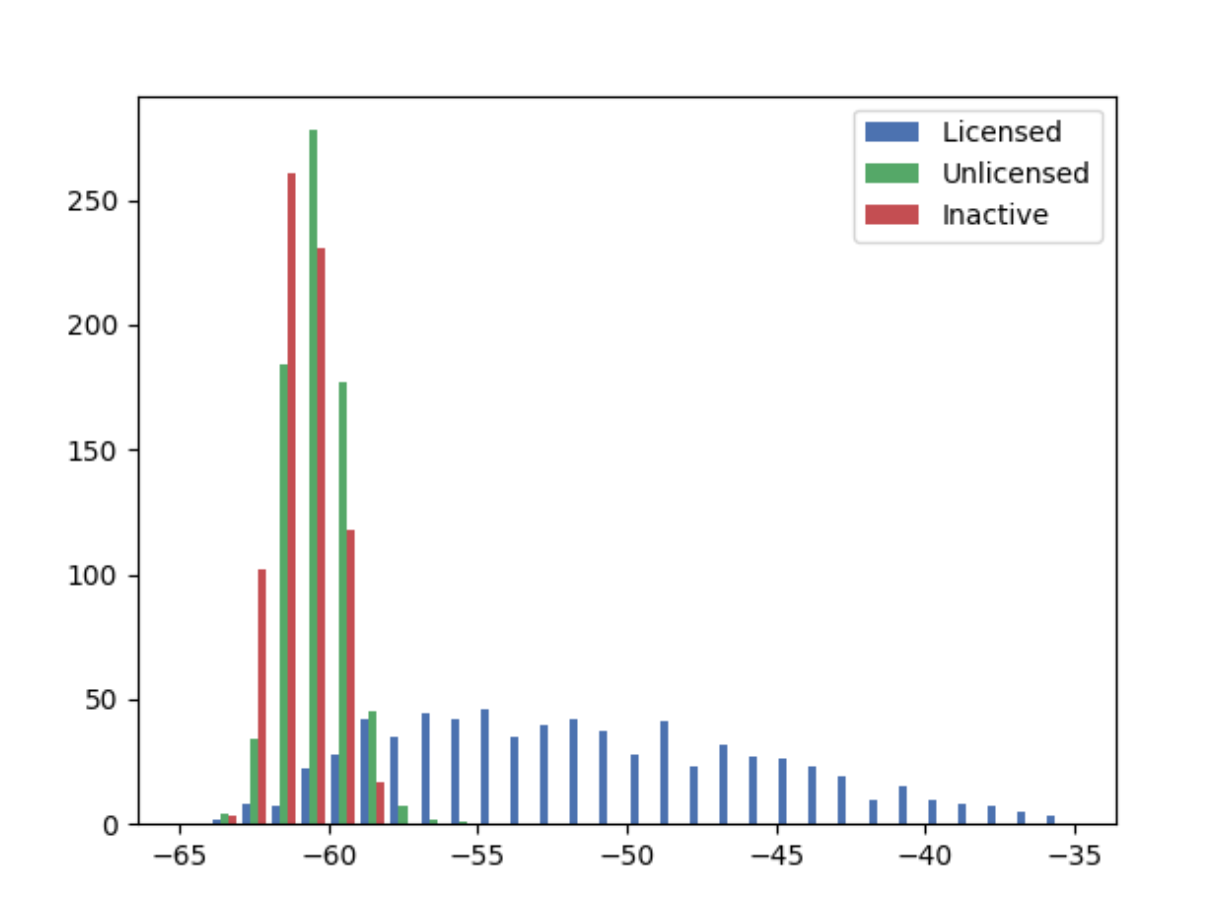}
  \end{minipage}
 \label{fig:new2}
 \caption{Comparison of Different Bands in the Same Scenario}
\end{figure}

 To analyze the difference between licensed and unlicensed bands, we plot the spectrogram of the data. Fig 3 is one set of data, including a complete test drive. Fig 3 (a) is the spectrogram of the measurement starting at 0 sec and ending at 950 sec. In this figure, different times also represent other locations. For licensed bands, we observe higher interference at locations closer to the cell tower (around 504 sec) than on the open road (0 – 216 sec). 1.9 GHz – 2.0 GHz has a similar spatial pattern as 2.1 – 2.2 GHz since the variations in power levels are almost the same. For the 2.4 GHz unlicensed band, the interference is discontinuous, and the power level is significantly lower than the licensed bands. Fig 3 (b) is the congested plot of the data showing the interference between 1.9 GHz and 2.5GHz. 2.0 GHz – 2.1 GHz and 2.2 -2.3 GHz seems to be inactive bands during the experiment, which shows the potential of spectrum sharing applications. 

\begin{figure*}[t!]
\centering
\hfill
\subfigure[]{
  \begin{minipage}[b]{0.3\textwidth}
    \centering
    \includegraphics[width=\textwidth]{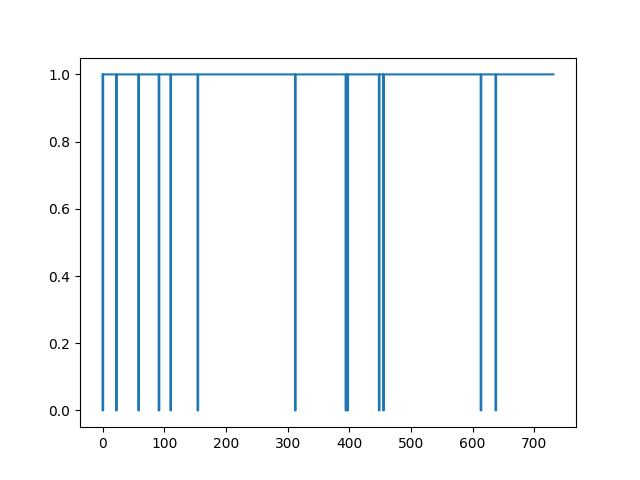}
    \label{fig:4_1}
  \end{minipage}
}
\hfill
\subfigure[]{
  \begin{minipage}[b]{0.3\textwidth}
    \centering
    \includegraphics[width=\textwidth]{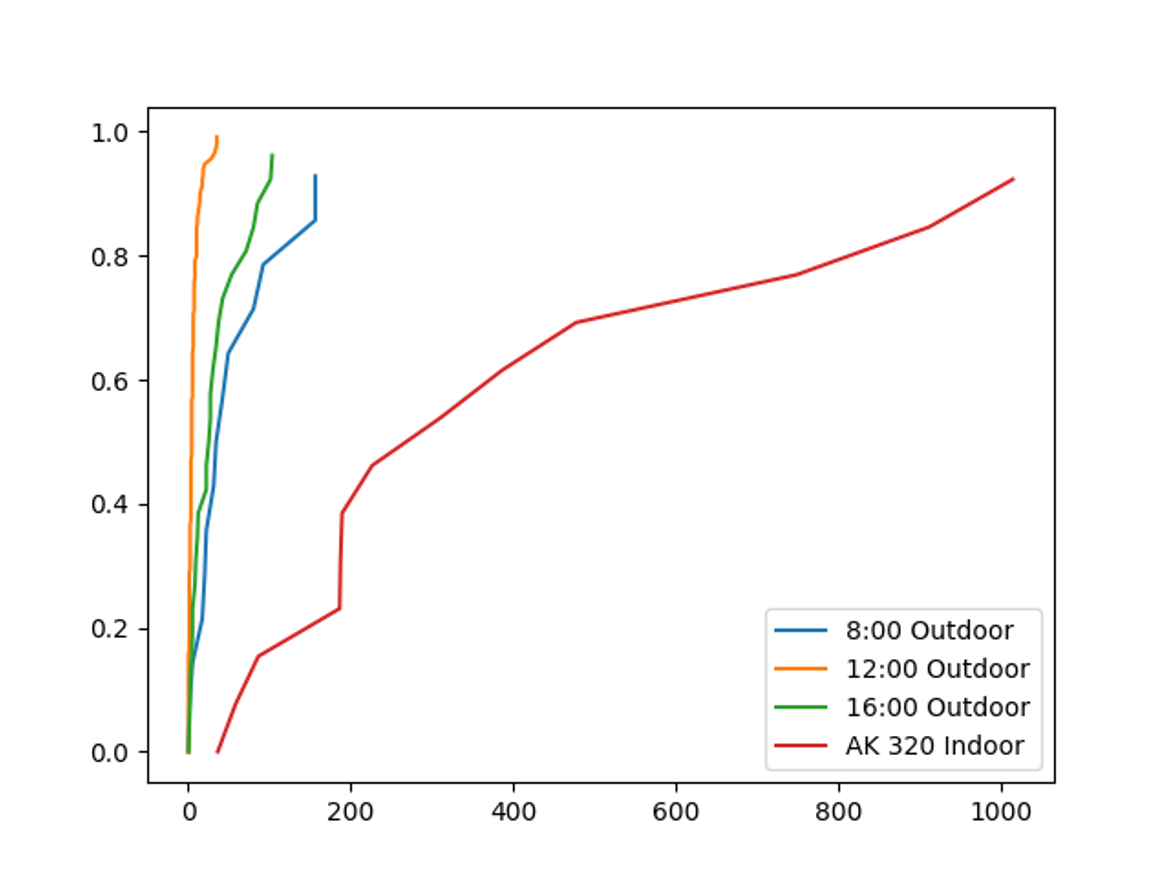}
    \label{fig:4_2}
  \end{minipage}
}
\hfill
\subfigure[]{
  \begin{minipage}[b]{0.3\textwidth}
    \centering
    \includegraphics[width=\textwidth]{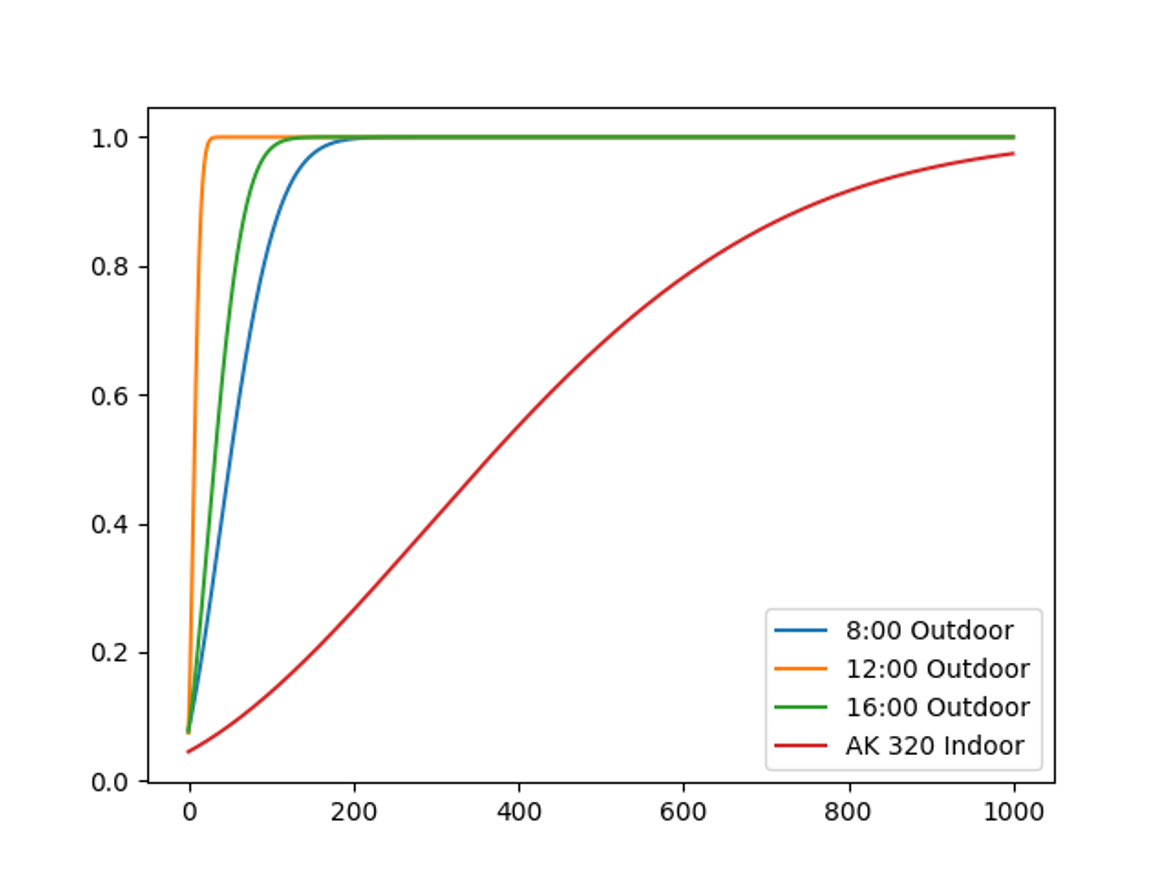}
    \label{fig:4_3}
  \end{minipage}
}

\caption{(a) Data Binarization (b) Empirical CDF (c) Fitted CDF}
\end{figure*}

\section{Interference Modeling and Comparison}
\label{sec:3}
In this section, we begin with the comparison using histograms. Fig. 4 reveals the difference in amplitude distribution of 4 datasets when we plot the histogram of the same frequency bin. Fig.\ref{fig:new_1} is the histogram of the licensed band. The indoor environment tends to have lower interference and is more centered. Fig.\ref{fig:new_2} and Fig.\ref{fig:new_3} are the plots for the unlicensed band and one of the unused bands, respectively. The main difference is the average interference level. We also compare the interference in each frequency band type by using only one dataset. Fig.\ref{fig:new2} shows histograms of 3 frequency bands in the same dataset. The variance of interference in the licensed band seems much more significant than in the other two bands. 
To have numerical comparisons, we can analyze the short-term variations of the interference in an IoT environment where many stationery and moving wireless devices interfere with each other based on circular scattering principles \cite{clarke1968statistical} \cite{pahlavan2005wireless}. This modeling enables a realistic performance evaluation of the RF cloud interference with any wireless device. According to Clarke’s Model, the probability density function of the amplitude fluctuations follows a Rayleigh Distribution: 	
\[f(r) = \frac{r}{{{\sigma ^2}}}{e^{ - \frac{{{r^2}}}{{2{\sigma ^2}}}}},r \ge 0\]
As a device moves along a path with velocity $v_m$, the Doppler shift from each interfering source depends on the spatial angle of the direction of movement with the direction of the source, $\alpha$,
\[{f_d} = \frac{{{v_m}}}{\lambda }\cos \alpha  = {f_m}\cos \alpha  \Rightarrow \alpha {\cos ^{ - 1}}\left( {\frac{{{f_d}}}{{{f_m}}}} \right)\]
For uniform distribution of interference angle, $\alpha$, the PDF ${f_A}(\alpha )$ is given as: 
\[{f_A}(\alpha ) = \frac{1}{{2\pi }};\alpha  \in ( - \pi ,\pi ]\]	 
The Doppler spectrum of the interference is \cite{pahlavan2005wireless}: 
 \begin{equation}
{\rm{D}}(f) = \frac{1}{{2\pi {f_m}}}{\left[ {1 - (f/{f_m})} \right]^{ - \frac{1}{2}}};\left| f \right| < {f_m}
\label{eq:doppler}
\end{equation}
Eq.\ref{eq:doppler} allows one to simulate the interference for a mobile user for performance analysis by running a complex Gaussian noise through a filter reflecting Doppler spectrum characteristics \cite{pahlavan2005wireless}.   Then we can design software and hardware interference simulators to examine the effects of IoT interference on a communication link, a GPS device, or a radar.    

\begin{figure*}[t!]
\centering
\hfill
\subfigure[]{
  \begin{minipage}[b]{0.3\textwidth}
    \centering
    \includegraphics[width=\textwidth]{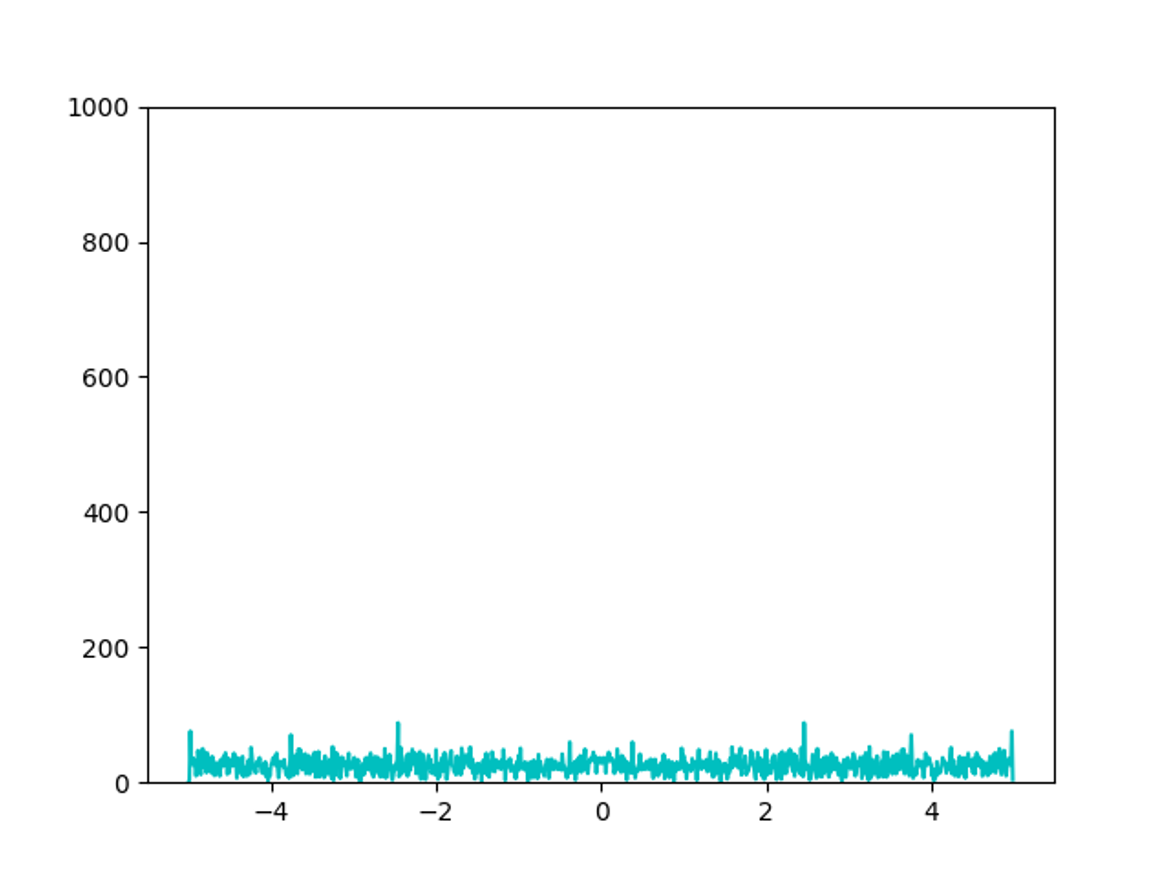}
    \label{fig:5_1}
  \end{minipage}
}
\hfill
\subfigure[]{
  \begin{minipage}[b]{0.3\textwidth}
    \centering
    \includegraphics[width=\textwidth]{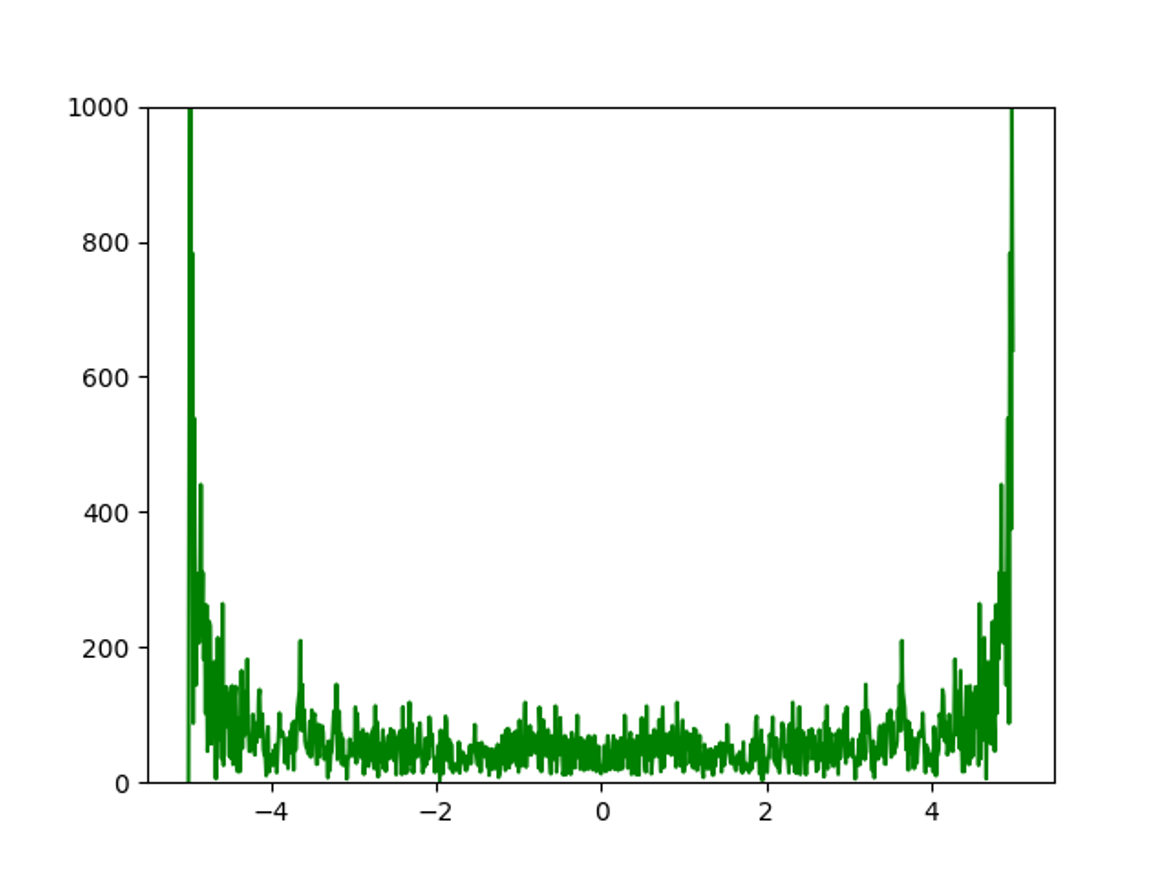}
    \label{fig:5_2}
  \end{minipage}
}
\hfill
\subfigure[]{
  \begin{minipage}[b]{0.3\textwidth}
    \centering
    \includegraphics[width=\textwidth]{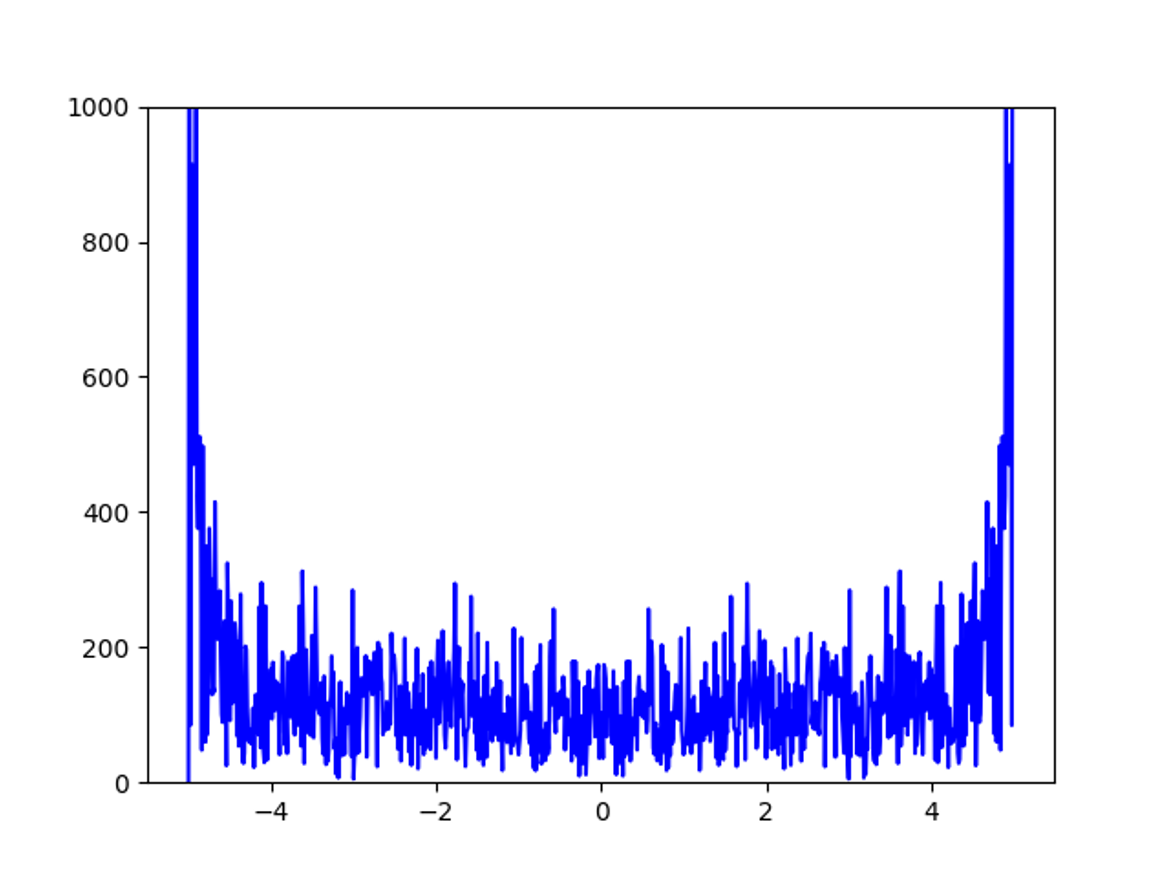}
    \label{fig:5_3}
  \end{minipage}
}
\caption{Doppler Spectrum (a) Inactive Band (2.2 GHz) (b) Unlicensed Band (2.4 GHz) (c) Licensed Band (1.95 GHz)}
\end{figure*}

For obtaining a model from empirical data, we first introduce data processing. The received power is first normalized using Min-Max scaling. The result normalized power is in the range [0, 1]. Then we convert the normalized power to a dummy variable using Eq.\ref{eq:1}, where $I$ is the interference reading, and $I_{new}$ donates the result dummy variable. The data binarization makes computing the spectrum occupancy much easier. 

\begin{equation}
{I_{new}} = \left\{ \begin{array}{l}
0,I \ge 0.1\\
1,I < 0.1
\end{array} \right.
\label{eq:1}
\end{equation}

We define the spectrum occupancy by counting the number of consecutive “1”s, which means if the current interference is greater than 10\% of the maximum interference in this channel, we consider the band as occupied and busy, and it is unavailable to spectrum sharing.
Fig. 4a shows the result of the binarization of one of the licensed frequency bands. We can only observe about ten “0”s, so this frequency band is highly occupied. If we count the number of  “1”s between every two “0”s, we can calculate the probability of the occupancy time for a specific frequency band. 
Using the approach described above, we calculate the Cumulative Distribution Function (CDF) using real data. Fig 4b shows the empirical CDF of 3 different times of the day for the outdoor environment and one indoor dataset. All three outdoor datasets almost reach a probability of 0.9 for occupancy<=200. Since the sample rate is 1 per second, it means we have a 90\% probability that the frequency occupancy is less than 200 seconds. For the indoor environment, the empirical CDF shows a probability of 0.22 for occupancy$<=$200. The outdoor environment tends to have less interference than the indoor environment because the indoor environment has more interference sources and a severe radio frequency (RF) environment. 
To numerical compare the effect of interfence on spectrum occupancy, we fit the empirical data to obtain distribution and then evaluate the parameter. Since  Rayleigh distribution is widely used for Wi-Fi channel occupancy modeling, we also apply it to the licensed band to have a fair evaluation. Fig 4c is the CDF of fitted Rayleigh distribution for four datasets mentioned before. Outdoor measurements approach 1 much faster than indoor measurements. The detailed parameters are shown in Table 1. 

\begin{table}[]
\caption{Rayleigh Distribution Parameters}
\label{table:1}
\centering
\begin{tabular}{|l|l|l|}
\hline
{Dataset} & {Loc} & {Scale} \\ \hline
8:00 Outdoor                   & -26.8908                   & 65.8010                      \\ \hline
12:00 Outdoor                  & -3.5453                    & 9.0201                       \\ \hline
16:00 Outdoor                  & -16.3339                   & 40.9457                      \\ \hline
AK 320 Indoor                  & 21.4714                    & 182.7133                     \\ \hline
\end{tabular}
\end{table}

We further investigate the frequency domain of interference measurements for indoor and outdoor scenarios. We first deduct the average value from the data using Eq.\ref{eq:2} to remove the DC component.

 \begin{equation}
{I_{new - i}} = {I_i} - \frac{1}{N}\sum\limits_i^N {{I_i}} 
\label{eq:2}
\end{equation}

Then we apply FFT to the processed data. Using one of the outdoor datasets as an example, Fig.\ref{fig:5_1} shows the frequency domain feature of an inactive band (2.2 GHz). Fig.\ref{fig:5_2} is the FFT result of the unlicensed band (2.4 GHz). Fig.\ref{fig:5_3} illustrates one of the licensed bands (1.95 GHz). Compared with the two more occupied bands, the inactive band shows fewer variances and tends to be “flat”. Both licensed and unlicensed bands have a U-shape spectrum. The only difference is that the unlicensed band has minor variations. Frequency domain patterns can also contribute to estimating whether the spectrum resource is available to be allocated to secondary users. This section analyzes the interference in two environments using statistical approaches and the interference in different frequency bands using frequency-domain approaches.

\section{CONCLUSION}
Spectrum monitoring using the spectrum analyzer provides an approach to read and log the interference in real-time. The collected dataset shows its potential for intelligence spectrum management with retrieved temporal and spatial information. In this study, we demonstrate that the interference is highly related to the location, time, and band type. By analyzing the probability distribution of the occupancy time, we can estimate how long we can allocate a currently unused frequency band to secondary users. The result in the frequency domain demonstrates different mechanisms should be considered for each type of frequency band while assigning frequency resources. The information obtained from spectrum monitoring is critical for intelligence spectrum management since it provides parameters for manege the interference. We believe more datasets and Deep Learning algorithms can make accurate dynamic spectrum allocation achievable.

\ifCLASSOPTIONcompsoc
  \section*{Acknowledgments}
I’m extremely grateful to my friends Jianan Li for driving me to collect the data and Fei Li for his help on pyVisa.
I’d also like to express my thanks to my classmates in ECE 538 for their patience in peer-reviewing this work.

\else
  \section*{Acknowledgment}
\fi

\ifCLASSOPTIONcaptionsoff
  \newpage
\fi


\nocite{*}
\bibliographystyle{IEEEtran}
\bibliography{citations}

\begin{thebibliography}{10}
\providecommand{\url}[1]{#1}
\csname url@samestyle\endcsname
\providecommand{\newblock}{\relax}
\providecommand{\bibinfo}[2]{#2}
\providecommand{\BIBentrySTDinterwordspacing}{\spaceskip=0pt\relax}
\providecommand{\BIBentryALTinterwordstretchfactor}{4}
\providecommand{\BIBentryALTinterwordspacing}{\spaceskip=\fontdimen2\font plus
\BIBentryALTinterwordstretchfactor\fontdimen3\font minus
  \fontdimen4\font\relax}
\providecommand{\BIBforeignlanguage}[2]{{%
\expandafter\ifx\csname l@#1\endcsname\relax
\typeout{** WARNING: IEEEtran.bst: No hyphenation pattern has been}%
\typeout{** loaded for the language `#1'. Using the pattern for}%
\typeout{** the default language instead.}%
\else
\language=\csname l@#1\endcsname
\fi
#2}}
\providecommand{\BIBdecl}{\relax}
\BIBdecl

\bibitem{pahlavan2021understanding}
K.~Pahlavan, ``Understanding of rf cloud interference measurement and
  modeling,'' \emph{International Journal of Wireless Information Networks},
  pp. 1--16, 2021.

\bibitem{news2022}
\BIBentryALTinterwordspacing
N.~Chokshi. (1999) The f.a.a. announces progress in expanding 5g service at
  airports. [Online]. Available:
  \url{https://www.nytimes.com/2022/01/28/technology/faa-5g-verizon-att.html}
\BIBentrySTDinterwordspacing

\bibitem{okegbile2021stochastic}
S.~D. Okegbile, B.~T. Maharaj, and A.~S. Alfa, ``Stochastic geometry approach
  towards interference management and control in cognitive radio network: A
  survey,'' \emph{Computer Communications}, vol. 166, pp. 174--195, 2021.

\bibitem{yun2021intelligent}
D.-W. Yun and W.-C. Lee, ``Intelligent dynamic spectrum resource management
  based on sensing data in space-time and frequency domain,'' \emph{Sensors},
  vol.~21, no.~16, p. 5261, 2021.

\bibitem{wang2016stochastic}
C.~Wang, C.~Yang, L.~Liu, P.~Wang, and H.~Liu, ``Stochastic geometry
  interference model for 5g heterogeneous network,'' in \emph{2016 17th
  International Conference on Parallel and Distributed Computing, Applications
  and Technologies (PDCAT)}.\hskip 1em plus 0.5em minus 0.4em\relax IEEE, 2016,
  pp. 286--289.

\bibitem{lu2021stochastic}
X.~Lu, M.~Salehi, M.~Haenggi, E.~Hossain, and H.~Jiang, ``Stochastic geometry
  analysis of spatial-temporal performance in wireless networks: A tutorial,''
  \emph{IEEE Communications Surveys \& Tutorials}, 2021.

\bibitem{hou2021modeling}
Y.~Hou, J.~Webber, K.~Yano, S.~Kawasaki, S.~Denno, and Y.~Suzuki, ``Modeling
  and predictability analysis on channel spectrum status over heavy wireless
  lan traffic environment,'' \emph{IEEE Access}, vol.~9, pp. 85\,795--85\,812,
  2021.

\bibitem{bhatti2021shared}
F.~A. Bhatti, M.~J. Khan, A.~Selim, and F.~Paisana, ``Shared spectrum
  monitoring using deep learning,'' \emph{IEEE Transactions on Cognitive
  Communications and Networking}, vol.~7, no.~4, pp. 1171--1185, 2021.

\bibitem{al2018free}
B.~Al~Homssi, A.~Al-Hourani, R.~J. Evans, K.~G. Chavez, S.~Kandeepan, W.~S.
  Rowe, and M.~Loney, ``Free spectrum for iot: How much can it take?'' in
  \emph{2018 IEEE International Conference on Communications Workshops (ICC
  Workshops)}.\hskip 1em plus 0.5em minus 0.4em\relax IEEE, 2018, pp. 1--6.

\bibitem{clarke1968statistical}
R.~H. Clarke, ``A statistical theory of mobile-radio reception,'' \emph{Bell
  system technical journal}, vol.~47, no.~6, pp. 957--1000, 1968.

\bibitem{pahlavan2005wireless}
K.~Pahlavan and A.~H. Levesque, \emph{Wireless information networks}.\hskip 1em
  plus 0.5em minus 0.4em\relax John Wiley \& Sons, 2005.

\bibitem{kim2020adjacent}
H.-K. Kim, Y.~Cho, and H.-S. Jo, ``Adjacent channel compatibility evaluation
  and interference mitigation technique between earth station in motion and
  imt-2020,'' \emph{IEEE Access}, vol.~8, pp. 213\,185--213\,205, 2020.

\end{thebibliography}

\clearpage




\end{document}